\documentclass[journal=jacsat,manuscript=article]{achemso}

\usepackage[T1]{fontenc}
\usepackage{amssymb}
\usepackage{bm}
\usepackage{amsmath}
\usepackage{booktabs}
\usepackage{dcolumn}
\usepackage{chemfig}
\usepackage{multirow}
\usepackage{makecell}
\usepackage{array}

\urldef{\platformurl}\url{https://huggingface.co/spaces/sobinalosious92/POLYMER-PROPERTY}

\author{Sobin Alosious}
\affiliation{Lucy Family Institute for Data and Society,
University of Notre Dame, Notre Dame, IN, 46556, USA}
\alsoaffiliation{Department of Aerospace and Mechanical Engineering, University of Notre Dame, Notre Dame, IN, 46556, USA}
\author{Guoyue Xu}
\affiliation{Department of Aerospace and Mechanical Engineering, University of Notre Dame, Notre Dame, IN, 46556, USA}

\author{Yuhan Liu}
\affiliation{Department of Chemistry and Biochemistry,
University of Notre Dame, Notre Dame, IN, 46556, USA}

\author{Renzheng Zhang}
\affiliation{Department of Aerospace and Mechanical Engineering, University of Notre Dame, Notre Dame, IN, 46556, USA}

\author{Meng Jiang}
\affiliation{Lucy Family Institute for Data and Society,
University of Notre Dame, Notre Dame, IN, 46556, USA}
\alsoaffiliation{Department of Computer Science and Engineering, University of Notre Dame, Notre Dame, IN, 46556, USA.}

\author{Tengfei Luo}
\affiliation{Lucy Family Institute for Data and Society,
University of Notre Dame, Notre Dame, IN, 46556, USA}
\alsoaffiliation{Department of Aerospace and Mechanical Engineering, University of Notre Dame, Notre Dame, IN, 46556, USA}
\alsoaffiliation{Department of Chemical and Biomolecular Engineering, University of Notre Dame, Notre Dame, IN, 46556, USA.}
\alsoaffiliation{Center for Sustainable Energy at Notre Dame (ND Energy),
University of Notre Dame, Notre Dame, IN, 46556, USA}
\email{tluo@nd.edu}

\title{PolyDiscover: an integrated trust-aware platform for multi-objective polymer discovery and simulation validation}

\abbreviations{ML, GNN, MD, DFT, LLM}
\keywords{polymer informatics, polymer discovery platform, multi-objective optimization, molecular dynamics workflow, machine learning}

\begin{document}

\begin{abstract}
Data-driven polymer discovery depends on capabilities that remain fragmented across disconnected tools: property databases, surrogate models, screening scripts, literature mining, and simulation workflows. Here we present PolyDiscover, an integrated polymer-discovery platform that unifies multi-source property lookup, graph-neural-network property prediction, trust-aware multi-objective screening, literature evidence extraction, generative candidate exploration, and automated handoff to molecular-dynamics validation within a single environment. Building on the data and models of our previously published ADEPT--PolyGraphMT workflow, the platform harmonizes experimental, molecular-dynamics, density-functional-theory, and group-contribution data covering 28 properties across approximately 13,000 real and one million virtual polymers. Its discovery engine couples feasibility filtering and exact Pareto ranking with a trust score that integrates similarity to known polymers, internal physical consistency, predictive uncertainty, and synthesizability, followed by diversity-aware shortlist selection. In two case studies targeting lightweight thermally stable structural materials and thermally stable electrical insulators, screened with a monomer HOMO--LUMO-gap proxy for electronic insulation, the workflow reduced a million-polymer search space to shortlists of 30 chemically diverse candidates. Higher-fidelity molecular-dynamics and density-functional-theory calculations on the shortlisted candidates, interpreted after correction for the known systematic density underestimation of the molecular-dynamics protocol, showed that they retain the targeted property combinations, and an ablation analysis showed that trust-aware ranking shifts about one fifth of each shortlist toward more synthesizable chemistry that is closer to known polymers. A constrained natural-language interface translated a free-text design request into a validated specification that reproduced expert-defined screening exactly, and staged literature evidence corroborated the high glass-transition temperatures of the polyimide-dominated shortlists against published experiments. PolyDiscover thus provides a traceable route from design intent to candidates checked by higher-fidelity simulation and lowers the barrier to systematic exploration of polymer chemical space.
\end{abstract}

\section{Introduction}
Polymers underpin a wide range of structural, electronic, membrane, dielectric, thermal-management, and energy technologies because their chemistry and morphology can be tuned across an unusually large design space.\cite{tran2024aipolymers,kuenneth2023polybert,alosious2025computation} That same flexibility makes polymer discovery intrinsically multi-objective: candidate materials are rarely selected on the basis of a single property, but instead must balance thermal, mechanical, transport, optical, and dielectric performance together with processability and synthetic practicality.\cite{chen2021criticalnextsteps,amamoto2022datadriven,tran2024aipolymers} As a result, polymer design is not constrained primarily by the lack of promising hypotheses, but by the difficulty of navigating an enormous chemical and structural search space while preserving physical realism and application relevance.\cite{audus2017polyinformatics,chen2021criticalnextsteps,kuenneth2023polybert}

These challenges are amplified by the distinctive character of polymer data. Polymer properties depend not only on repeat-unit chemistry but also on chain architecture, higher-order structure, composition, and processing history, all of which are reported unevenly across the literature and are not always represented in machine-readable form.\cite{audus2017polyinformatics,amamoto2022datadriven,hatakeyama2023experimentoriented} Reviews of polymer informatics have repeatedly identified heterogeneous reporting, limited standardized representations, sparse property coverage, and the difficulty of encoding experimental context as major obstacles to robust data-driven design.\cite{audus2017polyinformatics,chen2021criticalnextsteps,hatakeyama2023experimentoriented} At the same time, much potentially valuable quantitative evidence remains embedded in unstructured papers rather than reusable databases, motivating recent efforts to extract polymer-property information directly from the literature using natural-language-processing pipelines and agent-based data-collection frameworks.\cite{shetty2023polymercorpora,yang2025quest2data} High-fidelity validation further compounds the bottleneck, because molecular simulation and quantum-chemical workflows remain computationally intensive and often require stitching together multiple software packages and heterogeneous outputs.\cite{atwi2022mispr,chen2025xpb}

Over the past several years, polymer informatics has nevertheless advanced rapidly. Polymer Genome established an influential precedent for web-accessible polymer property prediction based on curated data and surrogate models.\cite{kim2018polymergenome} More recent learning systems have expanded the scale and scope of prediction through multitask graph neural networks (GNNs) and chemical language models, enabling faster screening across larger polymer spaces and broader property sets,\cite{gurnani2023multitaskgnn,kuenneth2023polybert} and community benchmarking efforts have begun to standardize the evaluation of multi-task polymer property prediction.\cite{liu2025openpolymer} Multi-objective optimization has likewise become an established strategy for polymer screening and design: Pareto-front construction over surrogate-predicted properties was demonstrated early for organic dielectric polymers,\cite{mannodi2016pareto} and evolutionary and AI-assisted multi-objective searches have since guided polymer design for energy-storage and thermal-management applications.\cite{kern2021design,gurnani2024aiassisted,liu2026active} In parallel, broader reviews have highlighted the emergence of inverse design, generative modeling, and AI-assisted search strategies as increasingly practical components of polymer development workflows.\cite{chen2021criticalnextsteps,tran2024aipolymers,liu2025machine}

Complementary progress has occurred in infrastructure for computation and user interaction. MISPR demonstrated how automated, provenance-aware workflow systems can couple density-functional-theory (DFT) and molecular-dynamics (MD) calculations within a database-backed multiscale environment.\cite{atwi2022mispr} RadonPy established an automated all-atom MD pipeline for polymer property calculation,\cite{hayashi2022radonpy} while PolyOmics demonstrated the construction of an omics-scale computational polymer database containing more than $10^5$ polymers generated through fully automated MD workflows.\cite{yoshida2025omics} XPB addressed another important bottleneck by enabling high-throughput and chemically realistic generation of complex polymer structures for downstream simulation.\cite{chen2025xpb}
 PEMD extended this line of work toward an application-focused platform for polymer electrolyte materials, while ToPolyAgent showed how natural-language interfaces and large-language-model (LLM) agents can reduce the operational barrier to polymer simulation workflows.\cite{tan2026pemd,ding2026topolyagent} These studies represent substantial progress toward more accessible polymer computation. They also reveal that current tools tend to emphasize one dominant layer of the discovery process, such as property prediction, structure construction, simulation execution, or language-mediated control, rather than a single public-facing workflow that connects evidence gathering, property lookup, candidate screening, and simulation handoff in one environment.\cite{kim2018polymergenome,atwi2022mispr,hayashi2022radonpy,chen2025xpb,tan2026pemd,ding2026topolyagent}

This work addresses that integration challenge by introducing PolyDiscover,\footnote{Publicly accessible at \url{https://huggingface.co/spaces/sobinalosious92/POLYMER-PROPERTY}.} a polymer-discovery framework that connects property lookup, machine-learning (ML) prediction, multi-objective screening, literature evidence gathering, AI-assisted specification generation, generative candidate exploration, and automated simulation handoff within a unified computational environment. The platform is designed not merely as a collection of predictive tools, but as a traceable discovery workflow that supports progression from candidate definition to higher-fidelity validation. To support large-scale exploration, the framework integrates approximately 13,000 experimentally grounded polymers together with approximately 1 million virtual polymers and organizes 28 thermal, mechanical, transport, electronic, optical, and structural properties under a common schema. The property data and predictive models underlying the platform originate from our previously published ADEPT--PolyGraphMT workflow;\cite{alosious2026adept} the present work contributes the integrated discovery environment built on those resources, including the trust-aware screening engine, the natural-language and literature modules, and the automated simulation handoff.

At the core of the framework is a trust-aware multi-objective discovery engine that combines feasibility filtering, exact Pareto ranking, trust-aware candidate prioritization, and diversity-aware selection. Candidate shortlists can be further evaluated through literature-supported evidence and exported directly into the ADEPT molecular-simulation workflow for higher-fidelity validation. The capabilities of the platform are demonstrated through representative polymer-design problems targeting lightweight-structural and dielectric-material applications, complemented by a demonstration that natural-language interaction exactly reproduces expert-defined screening. These case studies demonstrate how the trust-aware discovery workflow can efficiently navigate million-scale polymer search spaces while preserving chemical diversity, physical plausibility, and traceability throughout the discovery and validation process.

\section{Platform Overview}
\begin{figure}[h!]
  \centering
  \includegraphics[width=0.75\textwidth]{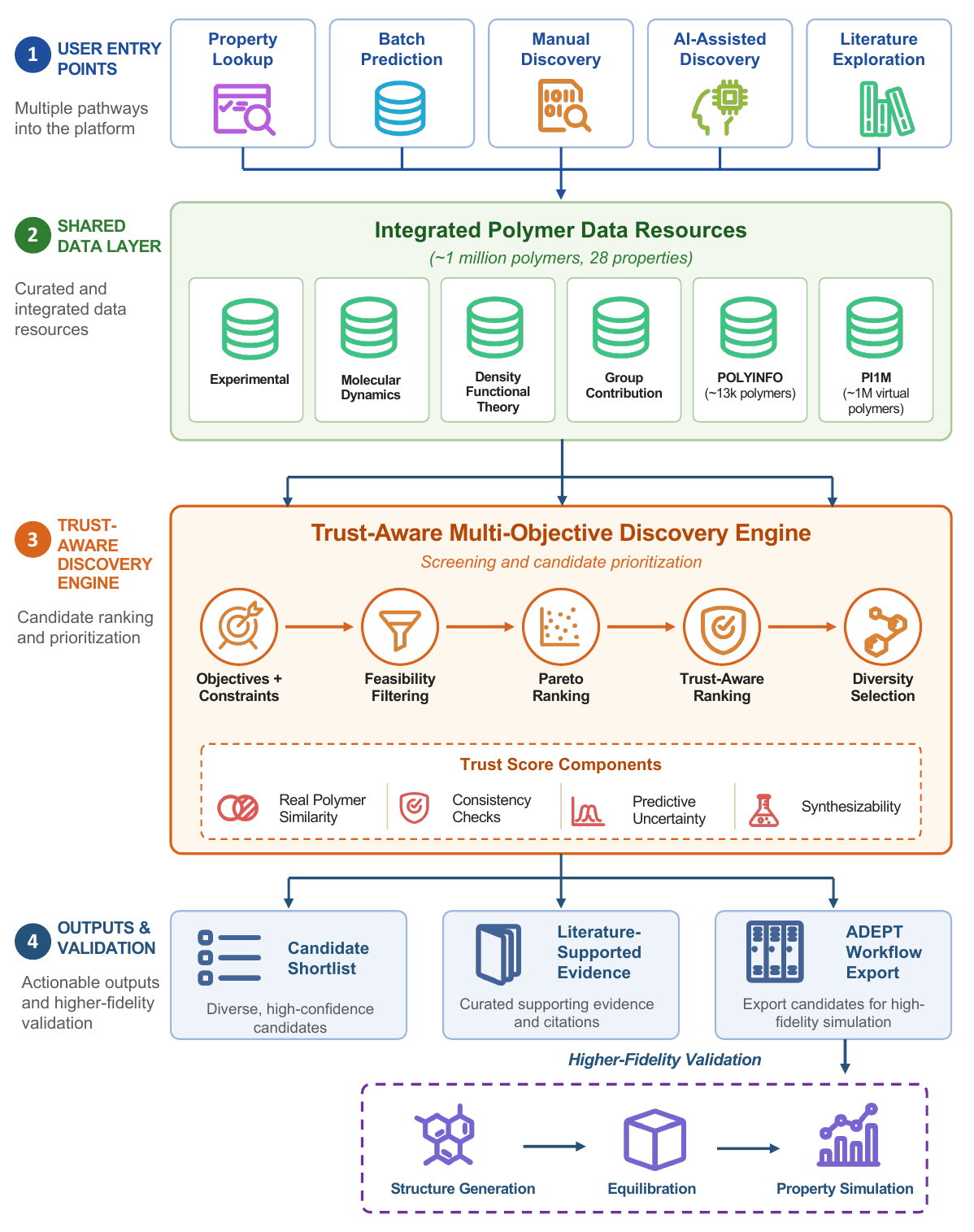}
\caption{
Overview of the PolyDiscover platform. Multiple user entry points, including property lookup, batch prediction, manual discovery, AI-assisted discovery, and literature exploration, access a shared data layer containing experimental, MD, DFT, group-contribution, POLYINFO, and PI1M resources spanning approximately one million polymers and 28 properties. Core analytical services support property lookup, property prediction, and literature evidence processing. Candidate prioritization is performed through a trust-aware multi-objective discovery engine that combines feasibility filtering, Pareto ranking, trust-aware ranking, and diversity-aware selection. The resulting candidate shortlists can be further evaluated through literature-supported evidence or exported to the ADEPT workflow for higher-fidelity simulation and validation.
}

\label{fig:platform_overview}
\end{figure}

PolyDiscover is designed as an integrated environment for polymer discovery and evaluation, combining data access, property prediction, multi-objective screening, literature-assisted evidence gathering, AI-assisted specification generation, and automated simulation handoff within a unified workflow. Rather than functioning as a collection of independent utilities, these components share a common data and decision framework that enables users to progress from candidate exploration to simulation-ready polymer systems while maintaining traceability across all stages of the workflow.

Users may enter the platform through several complementary pathways. Property lookup and batch prediction provide direct access to experimentally measured, simulation-derived, theory-derived, and ML-predicted polymer properties. The discovery interface supports both manually defined and natural-language-defined optimization problems, enabling users to formulate candidate-selection objectives through either structured inputs or free-text descriptions. Additional modules provide literature-centered exploration and generative design of novel polymer candidates. Regardless of the entry pathway, all workflows access the same underlying data resources and ultimately connect to a shared candidate-prioritization framework.

\begin{figure}[b!]
  \centering
  \includegraphics[width=1.0\textwidth]{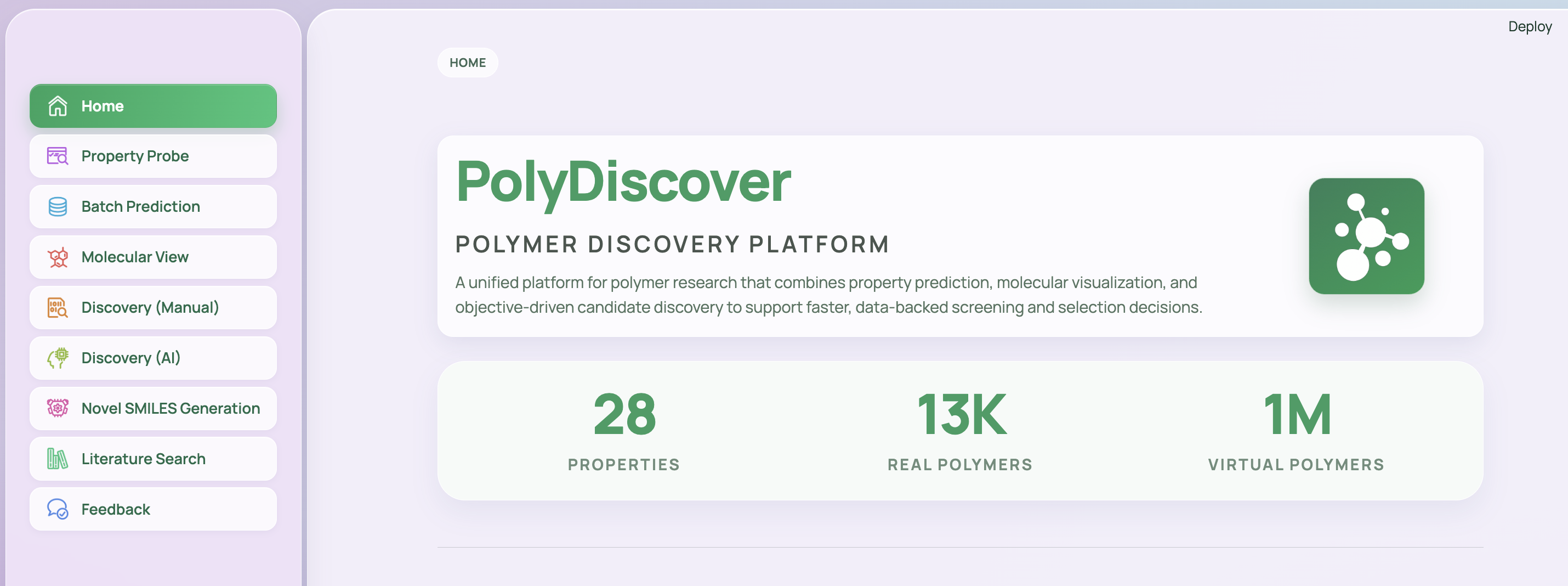}
\caption{
Home page of the publicly deployed PolyDiscover web application. The navigation panel provides access to the user-facing modules of the platform, including property lookup (Property Probe), batch prediction, molecular visualization, manual and AI-assisted discovery, novel SMILES generation, and literature search. The platform operates on approximately 13,000 real and one million virtual polymers and is publicly accessible at \platformurl.
}
\label{fig:platform_screenshot}
\end{figure}

Figure~\ref{fig:platform_overview} summarizes the overall organization of the platform. The integrated data layer combines experimental measurements, MD calculations, DFT calculations, group-contribution estimates, and large-scale real and virtual polymer libraries. These resources support both direct property retrieval and ML-based property completion. Candidate prioritization is performed through the discovery engine, after which shortlisted polymers can be further investigated through literature evidence gathering or exported to the ADEPT workflow for higher-fidelity simulation. All modules are exposed through a publicly deployed web application (Figure~\ref{fig:platform_screenshot}), so the complete workflow can be used from a standard web browser without local installation or programming expertise.

The current release integrates 13,000 experimentally grounded polymers from POLYINFO\cite{otsuka2011polyinfo} together with 1 million virtual polymers from the PI1M library \cite{ma2020pi1m}, resulting in a combined discovery space exceeding one million candidate polymers. Property information is organized using a unified schema spanning thermal, mechanical, transport, permeability, electronic, optical, and structural descriptors. Table~\ref{tab:property_inventory} lists the 28 properties covered by the platform together with their source-family availability and their participation in the prediction, screening, literature, and simulation-handoff modules. This common representation allows heterogeneous data sources to be accessed through a consistent interface while supporting screening, prediction, literature analysis, and simulation preparation within the same computational environment.

\begin{table*}[h!]
\centering
\scriptsize
\setlength{\tabcolsep}{4pt}
\begin{tabular}{lllcccccccc}
\toprule
Property & Symbol & Unit & EXP & MD & DFT & GC & Pred. & Screen & Lit. & ADEPT \\
\midrule
\multicolumn{11}{l}{\textit{Thermal}} \\
Melting temperature & $T_m$ & K & \checkmark &  &  &  & \checkmark & \checkmark & \checkmark &  \\
Glass transition temperature & $T_g$ & K & \checkmark & \checkmark &  &  & \checkmark & \checkmark & \checkmark & \checkmark \\
Thermal diffusivity & $\alpha_T$ & m$^2$/s &  & \checkmark &  &  & \checkmark & \checkmark & \checkmark & \checkmark \\
Thermal conductivity & $k$ & W m$^{-1}$ K$^{-1}$ & \checkmark & \checkmark &  &  & \checkmark & \checkmark & \checkmark & \checkmark \\
Specific heat capacity & $C_p$ & J kg$^{-1}$ K$^{-1}$ & \checkmark & \checkmark &  & \checkmark & \checkmark & \checkmark & \checkmark & \checkmark \\
\multicolumn{11}{l}{\textit{Mechanical}} \\
Young's modulus & $E$ & GPa &  & \checkmark &  &  & \checkmark & \checkmark & \checkmark & \checkmark \\
Shear modulus & $G$ & GPa &  & \checkmark &  &  & \checkmark & \checkmark & \checkmark & \checkmark \\
Bulk modulus & $K$ & GPa &  & \checkmark &  &  & \checkmark & \checkmark & \checkmark & \checkmark \\
Poisson ratio & $\nu$ & -- &  & \checkmark &  &  & \checkmark & \checkmark & \checkmark & \checkmark \\
\multicolumn{11}{l}{\textit{Transport and permeability}} \\
Viscosity & $\eta$ & Pa s &  & \checkmark &  &  & \checkmark & \checkmark & \checkmark & \checkmark \\
Diffusivity & $D$ & cm$^2$/s &  & \checkmark &  &  & \checkmark & \checkmark & \checkmark & \checkmark \\
He permeability & $P_{\mathrm{He}}$ & Barrer & \checkmark &  &  &  & \checkmark & \checkmark & \checkmark &  \\
H$_2$ permeability & $P_{\mathrm{H_2}}$ & Barrer & \checkmark &  &  &  & \checkmark & \checkmark & \checkmark &  \\
CO$_2$ permeability & $P_{\mathrm{CO_2}}$ & Barrer & \checkmark &  &  &  & \checkmark & \checkmark & \checkmark &  \\
N$_2$ permeability & $P_{\mathrm{N_2}}$ & Barrer & \checkmark &  &  &  & \checkmark & \checkmark & \checkmark &  \\
O$_2$ permeability & $P_{\mathrm{O_2}}$ & Barrer & \checkmark &  &  &  & \checkmark & \checkmark & \checkmark &  \\
CH$_4$ permeability & $P_{\mathrm{CH_4}}$ & Barrer & \checkmark &  &  &  & \checkmark & \checkmark & \checkmark &  \\
\multicolumn{11}{l}{\textit{Electronic and optical}} \\
Polarizability & $\alpha$ & a.u. &  &  & \checkmark &  & \checkmark & \checkmark & \checkmark & \checkmark \\
HOMO energy & HOMO & eV &  &  & \checkmark &  & \checkmark & \checkmark & \checkmark & \checkmark \\
LUMO energy & LUMO & eV &  &  & \checkmark &  & \checkmark & \checkmark & \checkmark & \checkmark \\
Monomer HOMO--LUMO gap & $E_g^{\mathrm{mono}}$ & eV &  &  & \checkmark &  & \checkmark & \checkmark & \checkmark & \checkmark \\
Dipole moment & $\mu$ & Debye &  &  & \checkmark &  & \checkmark & \checkmark & \checkmark & \checkmark \\
Total electronic energy & $E_{\mathrm{tot}}$ & eV &  &  & \checkmark &  & \checkmark & \checkmark & \checkmark & \checkmark \\
Refractive index & $n$ & -- &  & \checkmark &  &  & \checkmark & \checkmark & \checkmark & \checkmark \\
Dielectric constant & $\varepsilon_r$ & -- & \checkmark & \checkmark &  &  & \checkmark & \checkmark & \checkmark & \checkmark \\
Permittivity & $\varepsilon$ & -- &  & \checkmark &  &  & \checkmark & \checkmark & \checkmark & \checkmark \\
\multicolumn{11}{l}{\textit{Structural and physical}} \\
Radius of gyration & $R_g$ & \AA &  & \checkmark &  &  & \checkmark & \checkmark & \checkmark & \checkmark \\
Density & $\rho$ & g cm$^{-3}$ & \checkmark & \checkmark &  &  & \checkmark & \checkmark & \checkmark & \checkmark \\
\bottomrule
\end{tabular}
\caption{Property inventory and source coverage used by the platform. Source-family availability is shown for experimental measurements (EXP), molecular-dynamics calculations (MD), density-functional-theory calculations (DFT), and group-contribution estimates (GC), based on the shipped lookup tables. ``Pred.'' indicates that the property is available through the routed surrogate prediction layer. ``Screen'' indicates direct participation in multi-objective discovery on the precomputed real-plus-virtual screening tables. ``Lit.'' indicates that the property belongs to the standardized evidence schema used by the literature module. ``ADEPT'' indicates that the property can be interpreted directly within the downstream simulation-handoff workflow or its analysis outputs.}
\label{tab:property_inventory}
\end{table*}

\section{Methods}

The methodological core of the platform combines source harmonization,
surrogate prediction, multi-objective discovery, literature evidence
staging, AI-assisted specification generation, generative exploration,
and simulation handoff. The workflow is designed to preserve
provenance across these stages while maintaining a clear boundary
between data-driven prioritization and downstream physics-based
validation.

\subsection{Multi-Source Property Lookup and Harmonization}
All source tables are mapped onto a common schema at load time. Structures are canonicalized with RDKit~\cite{landrum2013rdkit}, property labels are harmonized, and records are indexed by canonical repeat-unit representation. This allows experimentally measured, simulation-derived, theory-derived, and group-contribution values to be queried through a shared interface while preserving source identity. The representation used at this stage is deliberately repeat-unit-centric rather than morphology-resolved; it is therefore suitable for large-scale screening, but it does not encode chain architecture or processing history explicitly. When a requested property is absent from the source tables, the workflow falls back to the predictive layer rather than terminating the query.

\subsection{Ensemble Graph-Based Property Prediction}
Polymer repeat units are represented as molecular graphs derived from RDKit-parsed structures and evaluated with a routed ensemble of GNNs. Most properties are handled by dedicated single-task predictors, whereas selected related properties are modeled through multitask branches. Each property is served by an ensemble of five independently seeded training runs (seeds 7, 42, 99, 137, and 2023). The precomputed screening tables used in all discovery runs report the mean and the population standard deviation of the five ensemble members as the prediction and its spread; for responsiveness, the interactive single-structure predictor of the web application currently evaluates a two-member subset of the same ensemble. The underlying model architectures, their training dataset of approximately 62,000 property values spanning 28 properties, and per-property accuracy benchmarks are documented in the companion ADEPT--PolyGraphMT work,\cite{alosious2026adept} which also generated the precomputed property tables for the POLYINFO and PI1M libraries used by the discovery engine. The mean prediction supplies property completion outside the measured subset of the library, while the ensemble spread is propagated into the downstream ranking workflow as a lightweight uncertainty signal. The predictive ecosystem is compatible with prior multi-fidelity and
transfer-learning strategies developed for polymer property prediction.\cite{alosious2026transfer} As in the lookup layer, this model family learns from repeat-unit structure rather than from explicit mesoscale morphology, so its role is prioritization rather than mechanistic replacement of experiment or simulation.

\subsection{Multi-Objective Screening and Candidate Ranking}
\begin{figure}[b!]
  \centering
  \includegraphics[width=1.0\textwidth]{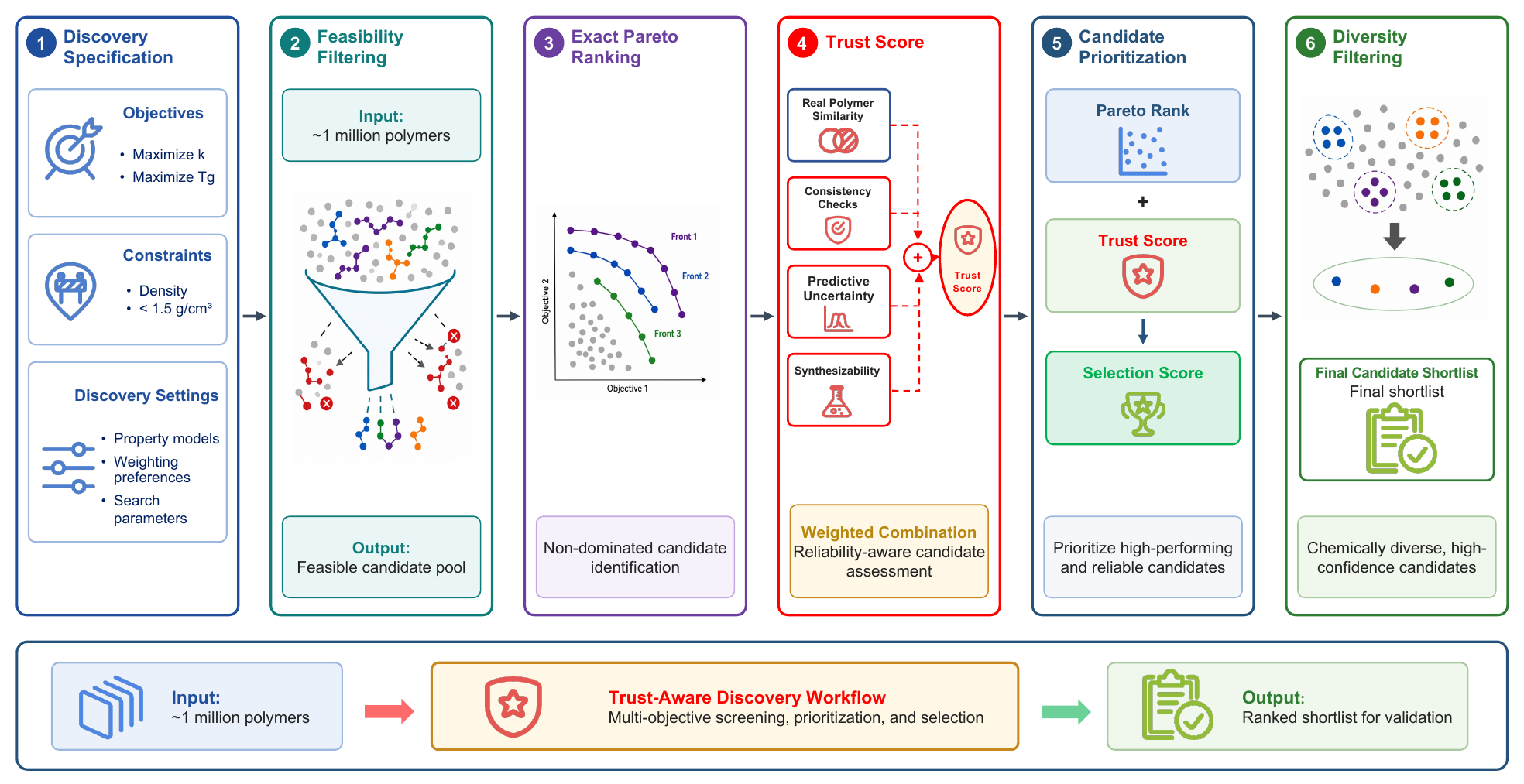}
\caption{
Trust-aware multi-objective discovery workflow implemented in the platform. User-defined objectives, constraints, and discovery settings are first applied through feasibility filtering to generate a feasible candidate pool from the polymer search space. Remaining candidates are ranked using exact Pareto-layer identification and subsequently evaluated using a trust score that combines similarity to known polymers, internal consistency checks, predictive uncertainty, and synthesizability. Pareto rank and trust score are integrated during candidate prioritization, after which diversity-aware filtering generates a chemically diverse, high-confidence shortlist for downstream literature analysis and simulation-based validation.
}

\label{fig:screening_workflow}
\end{figure}

Figure~\ref{fig:screening_workflow} summarizes the trust-aware discovery workflow, from feasibility filtering through Pareto ranking, trust scoring, and diversity-aware shortlist selection. 
The discovery workflow operates on precomputed screening tables and loads only the property columns required by the specified objectives and feasibility constraints. Candidates violating any hard constraint are discarded before ranking, and the remaining objective matrix is converted into a minimization problem where necessary. To support large candidate spaces, Pareto ranking is performed through exact chunked front extraction, yielding successive Pareto layers without requiring the entire library to be held in a single comparison pass. For interactive use, the engine can optionally subsample the feasible pool prior to Pareto ranking; this cap was disabled in all case studies reported here, so ranking was performed on the complete feasible set.

Candidate ranking is refined with a trust score designed as a risk-aware correction rather than a surrogate for physical truth. For each candidate $i$, the platform evaluates four quantities: proximity to known real polymers, an internal-consistency penalty, an uncertainty penalty, and a synthesizability proxy. The real-polymer term is computed from Morgan fingerprints (radius 2, 2048 bits) as
\begin{equation}
d_i^{\mathrm{real}} = 1 - \max_{j \in \mathcal{R}} \mathrm{Tan}\!\left(\mathbf{f}_i,\mathbf{f}_j\right), \qquad
s_i^{\mathrm{real}} = 1 - \mathrm{clip}\!\left(d_i^{\mathrm{real}},0,1\right),
\end{equation}
where $\mathcal{R}$ denotes the real-polymer reference set derived from the POLYINFO library, $\mathbf{f}_i$ is the fingerprint of candidate $i$, and $\mathrm{Tan}(\cdot,\cdot)$ is the Tanimoto similarity. The internal-consistency term penalizes violations of simple admissibility rules already enforced in the implementation, including positivity checks for selected properties (e.g., $\rho$, $C_p$, $k$, $D$, $\eta > 0$), non-negativity checks for $E_g^{\mathrm{mono}}$, $T_g$, and $T_m$, the bound $0 \leq \nu \leq 0.5$ for Poisson ratio, and the requirement $T_g \leq T_m$ when both values are present. If $N_i^{\mathrm{viol}}$ is the number of violated checks for candidate $i$ and $N_i^{\mathrm{chk}}$ is the number of checks that could be evaluated, then
\begin{equation}
p_i^{\mathrm{cons}} =
\begin{cases}
N_i^{\mathrm{viol}}/N_i^{\mathrm{chk}}, & N_i^{\mathrm{chk}} > 0,\\
0, & N_i^{\mathrm{chk}} = 0.
\end{cases}
\end{equation}
The uncertainty penalty is computed from the ensemble standard deviations after expressing each of them in units of the spread of the corresponding training targets, so that properties with different physical units contribute on a common scale,
\begin{equation}
\tilde{\sigma}_i = \frac{1}{|\mathcal{P}|}\sum_{p\in\mathcal{P}} \frac{\sigma_{i,p}}{s_p}, \qquad
p_i^{\mathrm{unc}} = \frac{\tilde{\sigma}_i}{\tilde{\sigma}_i + 1},
\end{equation}
where $\mathcal{P}$ is the set of properties involved in the active objectives and constraints, $\sigma_{i,p}$ is the ensemble standard deviation of property $p$ for candidate $i$, and $s_p$ is the standard deviation of the training targets of property $p$ stored with the trained models (112.5~K for $T_g$, 0.207~g~cm$^{-3}$ for $\rho$, and 1.56~eV for $E_g^{\mathrm{mono}}$). Without this normalization, the average would be dominated by the property with the largest numerical spread (typically $T_g$, expressed in kelvin), and the penalty would be nearly constant across candidates. Synthesizability is estimated from the RDKit synthetic-accessibility score and mapped linearly to a bounded score $s_i^{\mathrm{synth}} \in [0,1]$, with larger values indicating easier synthesis. These terms are combined as
\begin{equation}
S_i^{\mathrm{trust}} = w_{\mathrm{real}} s_i^{\mathrm{real}} + w_{\mathrm{cons}} (1-p_i^{\mathrm{cons}}) + w_{\mathrm{unc}} (1-p_i^{\mathrm{unc}}) + w_{\mathrm{synth}} s_i^{\mathrm{synth}},
\label{eq:trust_score}
\end{equation}
where the weights are normalized internally.
In the present work, the default platform weights
($w_{\mathrm{real}}=0.45$,
$w_{\mathrm{cons}}=0.25$,
$w_{\mathrm{unc}}=0.10$,
$w_{\mathrm{synth}}=0.20$)
were used for all discovery case studies.

Final candidate selection balances Pareto depth and trust score. If $L_i$ denotes the Pareto layer of candidate $i$ and $L_{\max}$ is the maximum number of retained Pareto layers ($L_{\max}=5$ in all case studies), with layers clipped to $[1, L_{\max}]$, the layer bonus is defined as
\begin{equation}
b_i^{\mathrm{pareto}} = \frac{(L_{\max}+1)-L_i}{L_{\max}},
\end{equation}
so that candidates in earlier Pareto layers receive larger bonuses. The overall selection score is then
\begin{equation}
S_i^{\mathrm{sel}} = w_{\mathrm{pareto}} b_i^{\mathrm{pareto}} + w_{\mathrm{trust}} S_i^{\mathrm{trust}}.
\label{eq:selection_score}
\end{equation}
Unless otherwise stated,
$w_{\mathrm{pareto}}=0.60$
and
$w_{\mathrm{trust}}=0.40$
were used throughout this work. The influence of the trust term on the final shortlists, and the sensitivity of the shortlists to all default weights, are quantified in the Results (Contribution of the Trust Score and Sensitivity to the Default Weights).

The final shortlist is selected greedily in descending order of $S_i^{\mathrm{sel}}$ under a molecular-diversity constraint,
\begin{equation}
1 - \mathrm{Tan}\!\left(\mathbf{f}_i,\mathbf{f}_j\right) \geq \delta_{\min},
\end{equation}
for all previously accepted candidates $j$, where $\delta_{\min}$ is the user-defined minimum Tanimoto distance ($\delta_{\min}=0.30$ in all case studies). Because the trust term is heuristic and partly model-derived, it should be interpreted as a ranking aid rather than as a calibrated probability of experimental success.

\subsection{Natural-Language Specification of Screening Objectives}

The natural-language interface is deliberately constrained. Rather than allowing arbitrary model-directed actions, the platform asks the language model to produce a structured discovery specification consisting of optimization objectives, hard feasibility constraints, and an intent profile selected from a fixed vocabulary (balanced, high-confidence, exploratory, high-diversity, strict-filtering, and real-only), optionally accompanied by up to two secondary profiles. Each profile corresponds to a documented preset adjustment of the candidate-selection settings: for example, the high-confidence profile shifts the selection weighting toward the trust score, whereas the balanced profile retains the platform defaults. All remaining run parameters are inherited from a fixed template rather than generated by the model. Returned specifications are sanitized before execution: unsupported properties are removed, malformed objectives and constraints are discarded or replaced by conservative defaults, and unrecognized or missing profile labels revert to the balanced profile. This design keeps the language layer auditable and ensures that candidate ranking remains entirely controlled by the deterministic screening engine.

\subsection{Literature Retrieval and Evidence Staging}
The literature module complements prediction and screening with source-grounded evidence. User queries can be framed around a material, a target property, or a broader research task, and are first processed by a rule-based clarifier that identifies opportunities for refinement without modifying the original request. The resulting search expression integrates the available material identifier, property specification, and free-text terms, and is executed deterministically across arXiv, PubMed, and Semantic Scholar. Retrieved records are aggregated and deduplicated by title, with each source contributing results in its native relevance order. A fixed number of records ($k$) from this pool is then passed to full-text acquisition. Records whose PDFs cannot be retrieved automatically are processed at the bibliographic-record level or supplemented through manual PDF upload. The archived query, parameters and retrieval outcomes provide a reproducible evidence base for downstream analysis.

Structured extraction is attempted in two stages. A contextual extraction backend is tried first when configured; if that stage returns no usable points, the workflow falls back to a simpler direct extraction pathway. Raw outputs are converted into a common evidence schema containing material identity, normalized property name, raw value, raw unit, experimental conditions, source quotation, and source location. Unit harmonization is then performed against the platform property catalog. For a property $p$ with raw value $v_i^{\mathrm{raw}}$ and reported unit $u_i^{\mathrm{raw}}$, the staged value is
\begin{equation}
\left(v_i^{\ast},u_p^{\ast}\right)=\mathcal{S}_p\!\left(v_i^{\mathrm{raw}},u_i^{\mathrm{raw}}\right),
\end{equation}
where $\mathcal{S}_p$ denotes the property-specific standardization map and $u_p^{\ast}$ is the platform-standard unit for property $p$. Values that cannot be parsed or converted are discarded before staging.

Each candidate evidence point is then validated against minimum structural requirements, including the presence of a material name, a normalized property key, a standardized numerical value, and a sufficiently informative source quotation. Property-aware plausibility bounds are additionally enforced for selected quantities. Valid evidence is assigned a quality tier from a simple additive score,
\begin{equation}
Q_i = 2I_i^{\mathrm{value}} + \min\!\left(N_i^{\mathrm{cond}},3\right) + I_i^{\mathrm{method}} + I_i^{\mathrm{loc}} + q_i^{\mathrm{conf}},
\end{equation}
where $I_i^{\mathrm{value}}$ indicates the presence of a standardized value, $N_i^{\mathrm{cond}}$ is the number of captured experimental-condition fields, $I_i^{\mathrm{method}}$ indicates an identified measurement method, $I_i^{\mathrm{loc}}$ indicates source localization, and $q_i^{\mathrm{conf}}$ contributes 2, 1, or 0 depending on whether the extraction confidence is at least 0.9, at least 0.7, or lower. In the current implementation, $Q_i \geq 7$ is labeled gold, $4 \leq Q_i < 7$ is labeled silver, and lower valid scores are labeled bronze. Because extraction quality varies across publishers and document formats, all staged evidence is routed through a reviewer queue, where each record can be approved, edited, or rejected; papers whose full text cannot be acquired automatically can be supplied through manual PDF upload, after which extraction is re-attempted.

\subsection{Generative Exploration of Novel Polymer Candidates}

Beyond screening existing polymers, the platform supports exploratory design through a generative workflow for proposing previously unseen repeat-unit structures. The implementation builds upon the PI1M framework, in which a recurrent neural network was trained on approximately 12,000 polymers collected from PolyInfo and subsequently used to generate a virtual library of approximately one million polymer repeat units.\cite{ma2020pi1m} The generative model operates on polymer-specific simplified molecular-input line-entry system (SMILES) representations that explicitly encode polymerization points, allowing the learned sequence model to capture characteristic polymer chemistry and generate chemically plausible repeat-unit structures. Generated candidates are canonicalized, deduplicated, and validated using RDKit-based chemical checks before being admitted into the searchable library. Additional novelty filtering is performed against the local reference datasets to distinguish newly generated structures from previously known polymers. The resulting workflow enables users to explore chemically valid and structurally diverse candidate polymers that extend beyond the experimentally reported chemical space while remaining compatible with the downstream prediction, screening, literature-evidence, and simulation-handoff modules of the platform.

\subsection{Automated Simulation Handoff}
The final stage of the workflow is the ADEPT simulation-handoff pathway.\cite{alosious2026adept} After shortlist generation, each selected polymer is assigned a persistent candidate identifier and written into a structure manifest that links identifiers to canonical repeat-unit strings. The exported archive includes this manifest together with job-control scripts, analysis utilities, and the directory structure expected by the downstream MD workflow. In the current implementation, the packaged assets are sufficient to launch array-based jobs in which each candidate identifier is mapped deterministically to an entry in the manifest and then propagated through the simulation stages.

The downstream branches exposed through the packaged workflow include amorphous-structure generation, equilibration, density and radius-of-gyration estimation, monomer-level electronic-property calculations, dielectric-property estimation, thermal-conductivity calculation, glass-transition and viscosity workflows, and equilibrium MD analysis of heat capacity, thermal expansivity, and elastic moduli. The role of the platform is not to execute those calculations directly or to replace the underlying physics models. Instead, it preserves candidate identity, structure provenance, and execution context so that the same shortlisted set can be transferred into simulation in a reproducible and auditable form.

\subsection{Higher-Fidelity Validation Calculations}
\label{sec:validation_methods}
All MD and electronic-structure validation results reported in this work were generated with the previously published ADEPT--PolyGraphMT workflow, and complete methodological details, convergence criteria, and benchmark comparisons are provided in that reference.\cite{alosious2026adept} Briefly, amorphous cells containing six chains of approximately 600 atoms each were built from repeat-unit SMILES strings at an initial density of 0.01~g~cm$^{-3}$, parameterized with the GAFF2 force field, and simulated with LAMMPS (Lennard-Jones cutoff 8~\AA, particle--particle particle--mesh electrostatics). After energy minimization, each cell was heated to 1000~K, compressed at 1000~K by ramping the pressure from 0.1 to 500~atm over 1~ns, relaxed at 1~atm, cooled from 1000 to 300~K over 5~ns at 1~atm, and equilibrated for a further 8~ns at 300~K and 1~atm in the isothermal--isobaric (NPT) ensemble; densities are averages over the final 2~ns of this run. As documented in the companion work,\cite{alosious2026adept} this protocol systematically underestimates experimental densities, so MD densities are compared with predictions after the linear calibration described in Case Study~1. Glass-transition temperatures were extracted from the temperature dependence of density using independently equilibrated simulations at discrete temperatures, with bilinear fits averaged over multiple fitting windows. Monomer HOMO--LUMO gaps ($E_g^{\mathrm{mono}} = E_{\mathrm{LUMO}} - E_{\mathrm{HOMO}}$) were computed for a single repeat unit whose two connection points are capped with (isotopically labelled) hydrogen atoms, using the range-separated, dispersion-corrected $\omega$B97M-D3BJ functional with the 6-311+G(2d,p) basis set. This is the quantity labelled ``band gap'' in the DFT source data; its relation to solid-state gaps is discussed in Case Study~2. These calculations follow the same protocols used to generate the MD and DFT source data of the platform, so screening and validation share a consistent physical foundation.

\section{Results}

\subsection{Application Case Studies}

The capabilities of the platform were evaluated through two polymer-discovery case studies representing lightweight-structural and electronic materials-design challenges, followed by a demonstration that the AI-assisted natural-language interface exactly reproduces the manually specified discovery workflow. Each case study was formulated as a constrained multi-objective optimization problem and executed on the merged discovery library of 1,009,140 entries (995,799 from PI1M and 13,341 from POLYINFO), Of these, 1,002,116 entries can be parsed by RDKit and correspond to 1,001,551 unique canonical repeat units, the remainder being repeated structures, often stored under different SMILES spellings; the other 7,024 entries cannot be parsed and are excluded at the diversity-selection stage. Repeated structures cannot enter the same shortlist twice, because their fingerprint (Tanimoto) distance is zero. Beyond identifying high-performing candidates, the aim was to assess whether the workflow could efficiently navigate large polymer search spaces and generate physically plausible candidates that remain favorable under higher-fidelity evaluation.

Unless otherwise stated, all studies employed the default discovery workflow consisting of feasibility filtering, Pareto ranking, trust-aware candidate prioritization, and diversity-aware shortlist generation.

\subsection{Case Study 1: Discovery of Lightweight Thermally Stable Polymers}
\label{sec:case1}

To demonstrate the practical application of the discovery workflow, we considered the design of lightweight polymers for applications requiring high thermal stability, such as aerospace structures, lightweight electronic packaging, and high-temperature engineering components. These applications demand materials that simultaneously possess low density for weight reduction and high glass-transition temperature to ensure dimensional and mechanical stability under elevated operating conditions.

The search was performed on the full merged discovery library (1,009,140 entries). Density ($\rho$) and glass-transition temperature ($T_g$) were selected as the optimization objectives:

\begin{equation}
\min(\rho), \qquad \max(T_g).
\end{equation}

This formulation seeks polymers that maximize thermal stability while minimizing material weight, two objectives that are often competing in practical polymer design.

The exact chunked Pareto-ranking procedure was applied directly to the complete discovery space. A total of 203 polymers were identified within the retained Pareto layers, from which a final shortlist of 30 chemically diverse candidates was selected using the trust-aware ranking and diversity-selection framework.

\begin{figure}[b!]
\centering
\includegraphics[width=\linewidth]{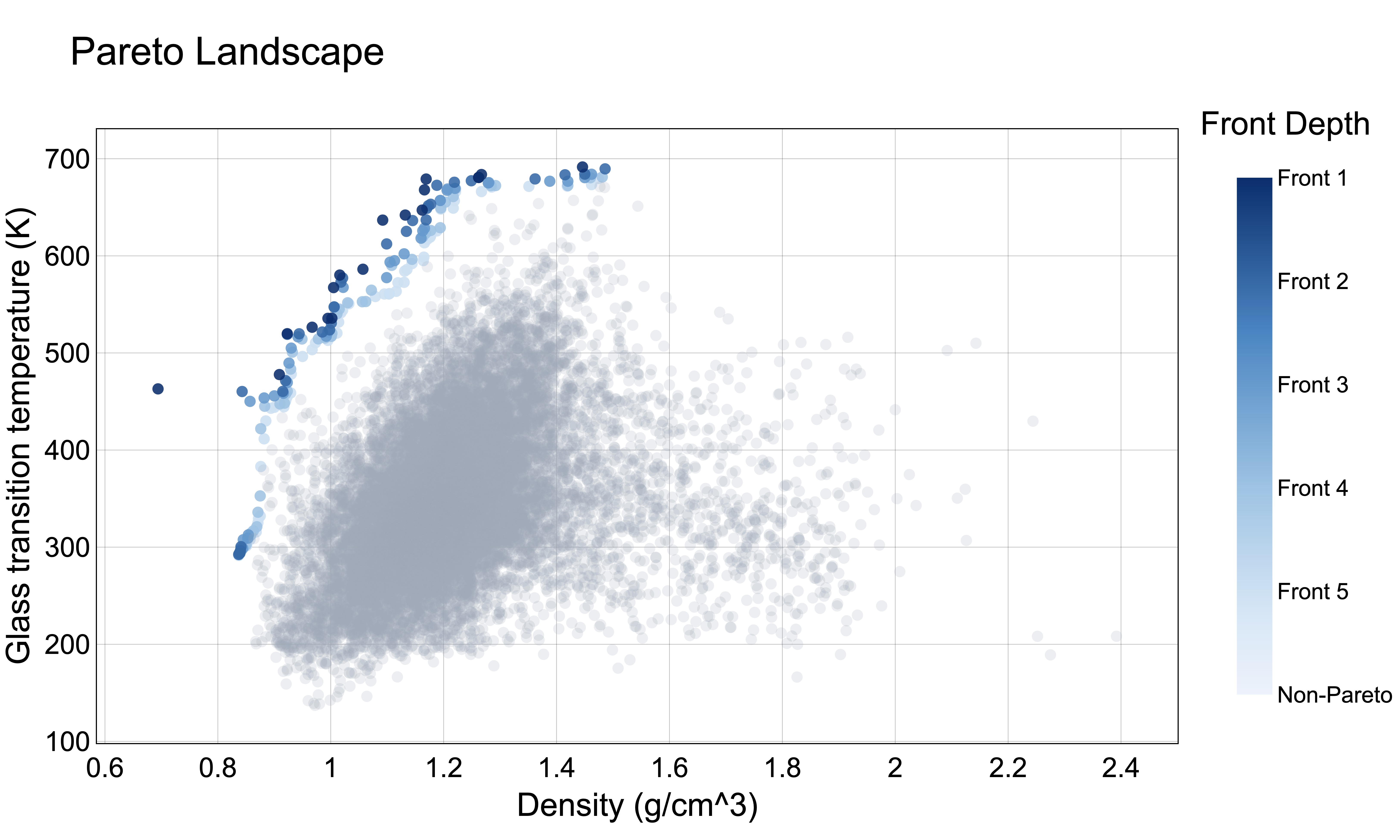}
\caption{
Pareto landscape for the lightweight thermally stable polymer discovery task. Density was minimized while glass-transition temperature was maximized over the merged discovery library containing 1,009,140 polymers. Gray points represent the screened polymer library, whereas colored points indicate polymers belonging to the retained Pareto layers. The final shortlist was obtained through trust-aware candidate prioritization and diversity-aware selection.
}
\label{fig:thermal_case_pareto}
\end{figure}

Figure~\ref{fig:thermal_case_pareto} presents the resulting Pareto landscape in the density--glass-transition-temperature objective space. Most polymers are concentrated in regions characterized by moderate densities and intermediate glass-transition temperatures, whereas the Pareto-optimal candidates define the non-dominated frontier toward simultaneously lower density and higher thermal stability. Rather than identifying a single optimum polymer, the workflow provides a diverse set of candidates representing different trade-offs between weight reduction and thermal performance, illustrating the advantage of multi-objective optimization over conventional single-property screening.

Shortlisted candidates were subsequently evaluated using the ADEPT MD workflow. The purpose of this validation step was not to calibrate the screening-stage predictions against MD calculations, but rather to determine whether highly ranked candidates retained favorable properties when assessed using a higher-fidelity physics-based simulation workflow.

\begin{table*}[t!]
\centering
\caption{ Higher-fidelity MD validation of five representative shortlisted candidates for the lightweight thermally stable polymer discovery task, ordered by trust score. Candidate ranking was performed by simultaneously minimizing density and maximizing glass-transition temperature. Trust scores use the normalized uncertainty penalty. ``MD (cal.)'' denotes MD densities corrected with the linear calibration against experiment ($\rho_{\mathrm{exp}} \approx 1.157\,\rho_{\mathrm{MD}} - 0.016$~g~cm$^{-3}$) fitted to 414 polymers. }
\label{tab:thermal_case_validation}
\scriptsize
\renewcommand{\arraystretch}{1.45}
\setlength{\tabcolsep}{5pt}

\begin{tabular}{>{\raggedright\arraybackslash}p{6.5cm}cccccc}
\toprule

\multirow{2}{*}{\textbf{Repeat Unit (SMILES)}} &
\multirow{2}{*}{\textbf{Trust Score}} &
\multicolumn{3}{c}{\textbf{Density (g cm$^{-3}$)}} &
\multicolumn{2}{c}{\textbf{$T_g$ (K)}} \\

\cmidrule(lr){3-5}
\cmidrule(lr){6-7}

&
&
Predicted &
MD &
MD (cal.) &
Predicted &
MD \\

\midrule

\makecell[l]{\ttfamily
*c1c(C)c(N2C(=O)c3ccc(-c4ccc5c(c4)C(=O)N(*)C5=O)\\
cc3C2=O)c(C)c2c1C(C)(C)CC2}
&
0.9353 &
1.2070 &
0.9826 &
1.121 &
668.85 &
650.56 \\

\makecell[l]{\ttfamily
*c1ccc(-n2c(=O)c3cc4c(=O)n(*)c(=O)\\
c4cc3c2=O)cc1}
&
0.9327 &
1.4860 &
1.2785 &
1.463 &
689.59 &
666.70 \\

\makecell[l]{\ttfamily
*C1C(=O)N(Nc2ccccc2)C(=O)C1*}
&
0.9018 &
0.8430 &
1.1487 &
1.313 &
460.29 &
525.75 \\

\makecell[l]{\ttfamily
*c1ccc(-c2cc(C)c(-n3c(=O)c4cc5c(=O)n(*)c(=O)\\
c5cc4c3=O)c(C)c2)cc1}
&
0.8269 &
1.3620 &
1.0918 &
1.247 &
679.25 &
759.27 \\

\makecell[l]{\ttfamily
*c1ccc2c(c1)C(=O)N(c1c(C)c(C)c(C)c\\
(N3C(=O)c4ccc(*)cc4C3=O)c1C)C2=O}
&
0.8069 &
1.2670 &
1.0481 &
1.196 &
683.79 &
661.87 \\

\bottomrule
\end{tabular}
\end{table*}

Table~\ref{tab:thermal_case_validation} summarizes five representative shortlisted candidates, selected for validation to span the chemistries represented in the shortlist while ensuring reliable completion of the automated simulation workflow. MD densities were obtained for 22 and MD glass-transition temperatures for 13 of the 30 shortlisted polymers, and the statistics below refer to all of them. Raw MD densities fall below the screening-stage predictions for 20 of the 22 polymers (median MD/predicted ratio 0.81). This offset is a property of the MD protocol rather than of the shortlist: across the 414 polymers for which the platform data contain both an ADEPT MD density and an experimental density, the MD value is lower in 97.6\% of cases (median MD/experiment ratio 0.88, interquartile range 0.85--0.91; median 0.90 for the four aromatic polyimides in this set), consistent with residual free volume in rapidly quenched, finite-chain amorphous cells.

A linear calibration fitted to these 414 polymers, following the procedure of the companion work,\cite{alosious2026adept} gives $\rho_{\mathrm{exp}} \approx 1.157\,\rho_{\mathrm{MD}} - 0.016$~g~cm$^{-3}$ and raises $R^2$ against experiment from 0.27 to 0.81, the calibrated value reported in the companion work (mean absolute error from 0.154 to 0.059~g~cm$^{-3}$). After calibration, the shortlisted MD densities lie at a median of 0.92 of the predicted values, and the four polyimides in Table~\ref{tab:thermal_case_validation} lie within 2--8\% of their predictions. MD glass-transition temperatures agree with the predictions with a mean absolute deviation of 59~K (Spearman rank correlation 0.70, $n = 13$), and all four polyimides in Table~\ref{tab:thermal_case_validation} exceed 650~K in MD.

The largest deviation occurs for poly[2,5-dioxo-1-(phenylamino)pyrrolidine-3,4-diyl], the third entry of Table~\ref{tab:thermal_case_validation} (19th of the 30 shortlisted candidates, Pareto layer~2; the table is ordered by trust score). It is predicted at 0.843~g~cm$^{-3}$ but simulated at 1.149~g~cm$^{-3}$ (1.313~g~cm$^{-3}$ after calibration). This polymer is a POLYINFO entry whose experimental density in the training data is 0.81~g~cm$^{-3}$, a value that is implausibly low for a polar imide bearing an aromatic substituent and that gives one of the three largest MD/experiment ratios among the 414 benchmark polymers. The surrogate reproduces this label: its five ensemble members split between 0.75--0.78 and 0.96~g~cm$^{-3}$, giving a standard deviation of 0.095~g~cm$^{-3}$, the largest density spread among the 203 Pareto-layer candidates. With the normalized uncertainty penalty it is the second most uncertain of these 203 candidates, but its exact match to a known polymer keeps its trust score high. The case shows that ensemble spread cannot reveal an erroneous training label and supports treating higher-fidelity validation as an integral stage of the workflow rather than an optional confirmation. Overall, the platform reduces a million-polymer search space to a small set of simulation-worthy candidates while preserving chemically diverse design alternatives.

\subsection{Case Study 2: Discovery of Thermally Stable Polymers with a Wide Monomer HOMO--LUMO Gap}
\label{sec:case2}

As a second demonstration, we considered the discovery of polymeric materials for electrically insulating and high-temperature electronic applications, including dielectric layers, encapsulation materials, and insulating structural components. Such applications require polymers possessing a large electronic band gap to suppress electronic conduction while maintaining high thermal stability during device operation. In addition, practical implementation often requires lightweight materials; therefore, the candidate density was constrained during screening.

The search was performed on the full merged discovery library (1,009,140 entries). The monomer HOMO--LUMO gap ($E_g^{\mathrm{mono}}$), used here as a proxy for electronic insulation, and the glass-transition temperature ($T_g$) were selected as the optimization objectives, while density was constrained to remain below 1.5 g cm$^{-3}$:

\begin{equation}
\max(E_g^{\mathrm{mono}}), \qquad \max(T_g)
\end{equation}

subject to

\begin{equation}
\rho \leq 1.5 \; \text{g cm}^{-3}.
\end{equation}

This formulation seeks polymers that simultaneously exhibit electrical insulation and thermal stability while satisfying practical lightweight design requirements.

The insulation objective requires care. $E_g^{\mathrm{mono}}$ is the HOMO--LUMO gap of a single hydrogen-capped repeat unit (see Methods), which is the quantity reported as the band gap in the DFT source data and predicted by the surrogate. Because it neglects conjugation along the chain, intermolecular interactions, and solid-state screening, it substantially overestimates solid-state gaps: across the 2,892 polymers of the DFT data it spans 5.5--16.8~eV (median 10.8~eV). Table~\ref{tab:gap_calibration} compares $E_g^{\mathrm{mono}}$ with the optical gaps of the eight polymers in the DFT data that Ohki et al.\ measured with a single method, extrapolation of the fundamental absorption edge recorded from the visible to the vacuum ultraviolet.\cite{ohki2010bandgap} The proxy exceeds the experimental values by 5.2--9.9~eV (median 7.0~eV), the largest offset occurring for PE, whose entry in the DFT data is a single CH$_2$ unit. It nevertheless largely preserves their ordering (Spearman rank correlation 0.88, Pearson correlation 0.94), with $E_g^{\mathrm{exp}} \approx 0.47\,E_g^{\mathrm{mono}} - 0.66$~eV. $E_g^{\mathrm{mono}}$ is therefore used only as a relative ranking descriptor for electronic insulation, and its absolute values should not be read as solid-state band gaps. This caveat is most important for aromatic polyimides, whose optical absorption is governed by intermolecular charge-transfer interactions that a single-repeat-unit calculation cannot capture. On the calibrated scale, PMDA--PDA (predicted $E_g^{\mathrm{mono}} = 8.3$~eV) corresponds to an optical gap of about 3.2~eV, and the Case Study 2 shortlist spans approximately 3.2--6.3~eV, with the PMDA-based polyimides at its low-gap end; their inclusion is driven by the $T_g$ objective. Kapton~H (PMDA--ODA), for which an optical gap of 2.2--2.3~eV has been reported,\cite{carotenuto2025kapton} lies below even this calibrated estimate, consistent with the additional charge-transfer absorption.

\begin{table}[t!]
\centering
\caption{Monomer HOMO--LUMO gap from the platform DFT data versus optical band gaps measured by Ohki et al.\ for eight commercial polymers with a single method (extrapolation of the fundamental absorption edge, visible to vacuum ultraviolet).\cite{ohki2010bandgap} The PE entry of the DFT data is a single CH$_2$ unit. Optical absorption edges lie below intrinsic electron--hole gaps; for PE, the photoconduction and photoemission threshold is 8.8~eV.\cite{less1973polyethylene}}
\label{tab:gap_calibration}
\small
\begin{tabular}{lcc}
\toprule
Polymer & $E_g^{\mathrm{mono}}$ (eV) & Optical gap (eV) \\
\midrule
Polyethylene (PE) & 16.78 & 6.9 \\
Polypropylene (PP) & 14.63 & 7.0 \\
Polycaprolactone (PCL) & 13.09 & 4.8 \\
Polyamide-6 (PA6) & 12.26 & 4.9 \\
Polylactide (PLA) & 11.39 & 5.0 \\
Polystyrene (PS) & 11.04 & 4.4 \\
Poly(ethylene terephthalate) (PET) & 9.89 & 4.0 \\
Poly(ethylene naphthalate) (PEN) & 8.53 & 3.3 \\
\bottomrule
\end{tabular}
\end{table}

Application of the density constraint retained 960,365 feasible polymers from the merged discovery space. The exact chunked Pareto-ranking procedure was subsequently applied to the feasible subset. A total of 261 polymers were identified within the retained Pareto layers, from which a final shortlist of 30 chemically diverse candidates was selected using the trust-aware ranking and diversity-selection framework.

\begin{figure}[b!] 
\centering \includegraphics[width=\linewidth]{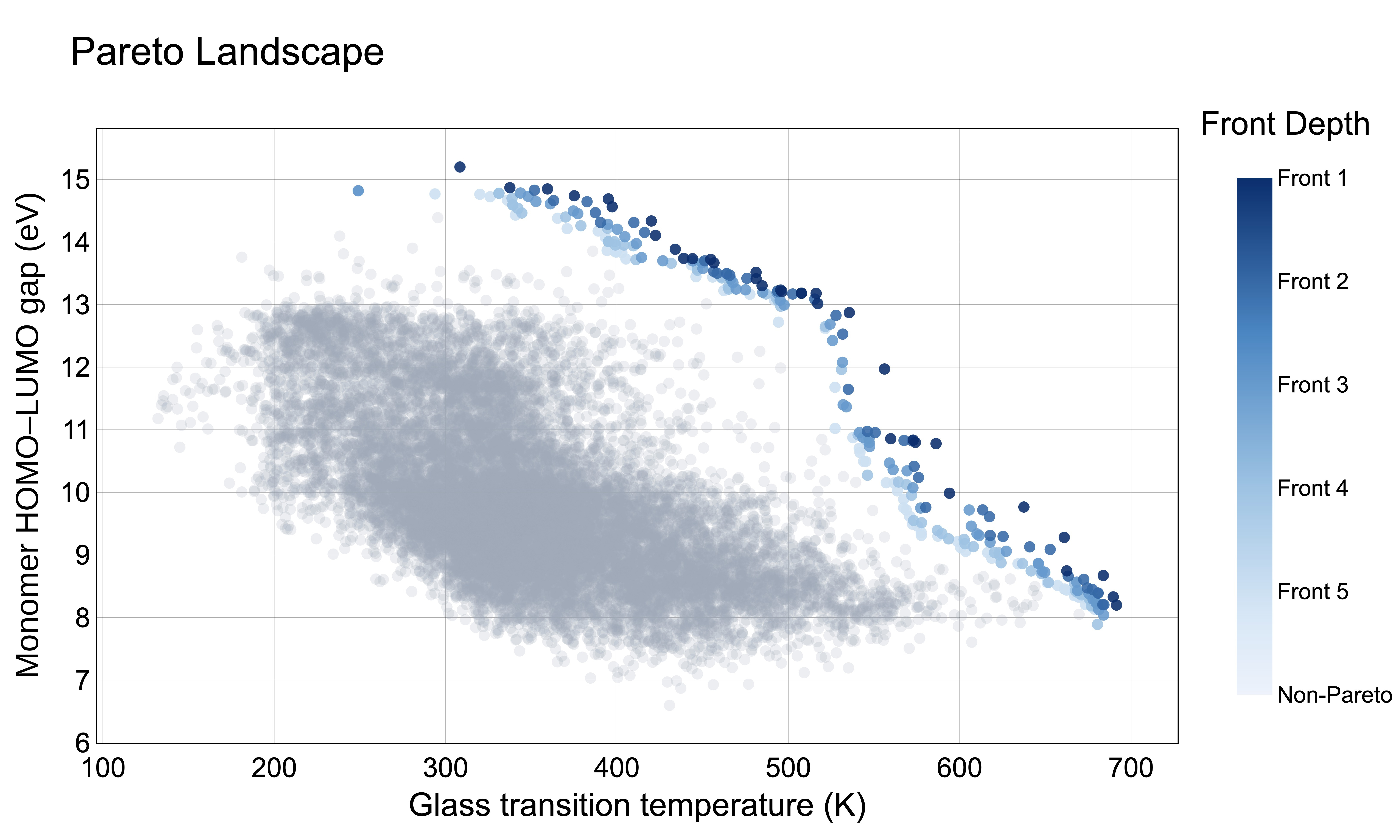} 
\caption{ Pareto landscape for Case Study 2. The monomer HOMO--LUMO gap ($E_g^{\mathrm{mono}}$) and glass-transition temperature were simultaneously maximized while constraining density to be below 1.5 g cm$^{-3}$. Gray points represent feasible polymers satisfying the density constraint, whereas colored points correspond to polymers belonging to the retained Pareto layers. The final shortlist was obtained through trust-aware candidate prioritization and diversity-aware selection. } 
\label{fig:bandgap_case_pareto} 
\end{figure}

Figure~\ref{fig:bandgap_case_pareto} presents the resulting Pareto landscape in the $E_g^{\mathrm{mono}}$--$T_g$ objective space. Most polymers populate the intermediate region characterized by moderate $E_g^{\mathrm{mono}}$ and glass-transition temperatures, whereas the Pareto-optimal candidates define the non-dominated frontier extending toward simultaneously larger $E_g^{\mathrm{mono}}$ and higher thermal stability. As expected for a multi-objective optimization problem, no single polymer dominates the entire search space. Instead, the workflow identifies a diverse set of candidates representing different trade-offs between electronic insulation and thermal robustness, allowing users to select materials according to application-specific priorities.

Shortlisted candidates were subsequently subjected to higher-fidelity validation. $E_g^{\mathrm{mono}}$ was recalculated with the DFT protocol described in the Methods, while density and glass-transition temperature were evaluated through the ADEPT MD workflow. The objective of this validation was not to establish one-to-one agreement with the screening-stage predictions, but rather to verify that the shortlisted candidates retained favorable electronic and thermomechanical characteristics under more rigorous computational evaluation.
\begin{table*}[t!]
\centering
\caption{
Higher-fidelity validation of five representative shortlisted candidates for Case Study 2, ordered by trust score. Candidate ranking simultaneously maximized the monomer HOMO--LUMO gap ($E_g^{\mathrm{mono}}$) and glass-transition temperature while satisfying the density constraint ($\rho \leq 1.5$ g cm$^{-3}$). $E_g^{\mathrm{mono}}$ was recalculated with DFT, and density and glass-transition temperature with MD; ``MD (cal.)'' denotes calibrated MD densities as in Table~\ref{tab:thermal_case_validation}. Trust scores use the normalized uncertainty penalty.
}
\label{tab:bandgap_case_validation}

\scriptsize
\renewcommand{\arraystretch}{1.45}
\setlength{\tabcolsep}{5pt}

\setlength{\tabcolsep}{2.5pt}
\begin{tabular}{>{\raggedright\arraybackslash}p{6.5cm}cccccccc}
\toprule

\multirow{2}{*}{\textbf{Repeat Unit (SMILES)}} &
\multirow{2}{*}{\textbf{Trust Score}} &
\multicolumn{2}{c}{\textbf{$E_g^{\mathrm{mono}}$ (eV)}} &
\multicolumn{3}{c}{\textbf{Density (g cm$^{-3}$)}} &
\multicolumn{2}{c}{\textbf{$T_g$ (K)}} \\

\cmidrule(lr){3-4}
\cmidrule(lr){5-7}
\cmidrule(lr){8-9}

&
&
Predicted &
DFT &
Predicted &
MD &
MD (cal.) &
Predicted &
MD \\

\midrule

\makecell[l]{\ttfamily
CC1=CC(C)=C(*)C(C)=C1N1C(=O)C2=CC=C(C=C2C1=O)\\
C1=CC2=C(C=C1)C(=O)N(*)C2=O}
&
0.9413 &
8.391 &
8.2537 &
1.262 &
1.0375 &
1.184 &
680.74 &
740.71 \\

\makecell[l]{\ttfamily
*c1ccc(-n2c(=O)c3cc4c(=O)n(*)c(=O)\\
c4cc3c2=O)cc1}
&
0.9337 &
8.329 &
8.0834 &
1.486 &
1.2824 &
1.467 &
689.59 &
702.06 \\

\makecell[l]{\ttfamily
*C1CC2CC1C(*)C2CCCC}
&
0.8735 &
13.015 &
13.0429 &
0.992 &
0.7864 &
0.893 &
517.17 &
513.79 \\

\makecell[l]{\ttfamily
*C1C(*)C2CC1C(C(=O)OCC13CC4CC(CC(C4)C1)C3)\\
C2C(=O)OCC12CC3CC(CC(C3)C1)C2}
&
0.8633 &
11.970 &
12.1677 &
1.136 &
0.9504 &
1.083 &
556.14 &
573.61 \\

\makecell[l]{\ttfamily
*N1C(=O)C2C3C=CC(C2C1=O)C1C(=O)N(\\
C24CC5CC(C2)C(C2C6CC7CC2CC(*)(C7)C6)\\
C(C5)C4)C(=O)C31}
&
0.8395 &
10.798 &
10.9357 &
1.182 &
1.0025 &
1.144 &
574.13 &
640.61 \\

\bottomrule
\end{tabular}
\end{table*}

Table~\ref{tab:bandgap_case_validation} summarizes five representative shortlisted candidates, selected on the same basis as in Case Study 1; DFT values were obtained for 29, MD densities for 28, and MD glass-transition temperatures for 18 of the 30 shortlisted polymers. The DFT recalculations reproduce the predicted $E_g^{\mathrm{mono}}$ closely (mean absolute deviation 0.22~eV, Spearman rank correlation 0.98, $n = 29$). Because the recalculation uses the same protocol as the DFT training data, and 7 of the 29 polymers are themselves part of those data, this comparison verifies surrogate fidelity rather than providing independent physical evidence of electronic insulation, whose interpretation is limited as discussed above. Raw MD densities are again below the predictions for all 28 polymers (median ratio 0.83). After the calibration described in Case Study~1, they lie at a median of 0.95 of the predicted values (range 0.88--1.03; mean absolute deviation 0.05~g~cm$^{-3}$), and all 28 satisfy the constraint $\rho \leq 1.5$~g~cm$^{-3}$ (maximum 1.48~g~cm$^{-3}$). MD glass-transition temperatures agree with the predictions with a mean absolute deviation of 54~K (Spearman rank correlation 0.92, $n = 18$). PMDA--PDA, which appears in both shortlists and was simulated independently in each case study, gives an indication of run-to-run variability: the two runs agree to within 0.3\% in density (1.278 and 1.282~g~cm$^{-3}$) and differ by 35~K in $T_g$ (666.7 and 702.1~K). Within the stated limitations of the gap proxy, these results show that the workflow identifies chemistries that combine a high $T_g$ and a wide $E_g^{\mathrm{mono}}$ while satisfying the density constraint.

\subsection{Contribution of the Trust Score and Sensitivity to the Default Weights}
\label{sec:trust_ablation}

With the default selection weights, one Pareto layer is worth 0.12 in $S_i^{\mathrm{sel}}$ (Eq.~\ref{eq:selection_score}). Across the Pareto-layer pools, the trust score varies by 0.41 (Case Study 1, 203 candidates) and 0.47 (Case Study 2, 261 candidates), which is worth up to 0.16--0.19 in $S_i^{\mathrm{sel}}$, so the trust term can reorder candidates across adjacent layers as well as within a layer. Because the diversity constraint is applied greedily in descending order of $S_i^{\mathrm{sel}}$, a reordering can also change which structures are admitted. To quantify the effect, both case studies were rerun from their Pareto-layer pools with the trust term removed ($w_{\mathrm{trust}} = 0$, Pareto rank plus diversity only, with 200 random tie-breaks within layers), with different selection weights, and with modified trust weights (Table~\ref{tab:trust_ablation}). Overlaps are computed on RDKit-canonical structures.

\begin{table*}[t!]
\centering
\caption{Ablation of the trust term and sensitivity of the shortlists to the default weights. Overlap is the fraction of the default 30-candidate shortlist that is retained (mean, with the minimum in parentheses for randomized variants). SA is the mean RDKit synthetic-accessibility score (lower is easier), $\bar{s}^{\mathrm{real}}$ the mean similarity to the nearest known polymer, and $\bar{L}$ the mean Pareto layer of the shortlist. Values are given as Case Study 1 / Case Study 2.}
\label{tab:trust_ablation}
\scriptsize
\renewcommand{\arraystretch}{1.25}
\begin{tabular}{>{\raggedright\arraybackslash}p{6.3cm}cccc}
\toprule
Variant & Overlap & SA & $\bar{s}^{\mathrm{real}}$ & $\bar{L}$ \\
\midrule
Default ($w_{\mathrm{pareto}}/w_{\mathrm{trust}} = 0.6/0.4$; trust weights 0.45/0.25/0.10/0.20) & 1.00 / 1.00 & 4.02 / 5.48 & 0.69 / 0.75 & 1.53 / 1.07 \\
Pareto rank + diversity only ($w_{\mathrm{trust}} = 0$; 200 tie-breaks) & 0.78 (0.67) / 0.82 (0.80) & 4.18 / 5.75 & 0.65 / 0.66 & 1.50 / 1.00 \\
$w_{\mathrm{trust}} = 0.2$ & 0.97 / 0.93 & 4.05 / 5.62 & 0.67 / 0.71 & 1.50 / 1.00 \\
$w_{\mathrm{trust}} = 0.6$ & 0.87 / 0.73 & 4.18 / 5.18 & 0.77 / 0.87 & 1.67 / 1.37 \\
$w_{\mathrm{trust}} = 0.8$ & 0.50 / 0.50 & 4.06 / 5.19 & 0.95 / 0.97 & 2.77 / 1.90 \\
Without real-polymer similarity & 0.93 / 0.83 & 4.03 / 5.53 & 0.66 / 0.67 & 1.50 / 1.00 \\
Without internal consistency & 0.93 / 0.93 & 4.05 / 5.43 & 0.73 / 0.79 & 1.60 / 1.13 \\
Without uncertainty & 0.97 / 0.97 & 4.00 / 5.45 & 0.71 / 0.77 & 1.57 / 1.10 \\
Without synthesizability & 0.87 / 0.90 & 4.15 / 5.55 & 0.75 / 0.81 & 1.63 / 1.17 \\
Trust weights, Dirichlet around defaults (200 draws) & 0.95 (0.83) / 0.96 (0.87) & 4.06 / 5.51 & 0.70 / 0.75 & 1.55 / 1.07 \\
Trust weights, uniform Dirichlet (200 draws) & 0.91 (0.83) / 0.91 (0.77) & 4.09 / 5.54 & 0.68 / 0.72 & 1.52 / 1.03 \\
\bottomrule
\end{tabular}
\end{table*}

Removing the trust term changes 22\% (Case Study 1) and 18\% (Case Study 2) of the shortlist on average, and up to 33\% and 20\%. Compared with the Pareto-plus-diversity selection, the trust-aware shortlists are more similar to known polymers (mean similarity 0.69 versus 0.65 and 0.75 versus 0.66), contain more polymers that are themselves in POLYINFO (37\% versus 31\% and 43\% versus 33\%), and have lower synthetic-accessibility scores (4.02 versus 4.18 and 5.48 versus 5.75), while remaining at essentially the same Pareto depth (mean layer 1.53 versus 1.50 and 1.07 versus 1.00). Without the trust term, only 3.4 and 3.6 of the five candidates in Tables~\ref{tab:thermal_case_validation} and~\ref{tab:bandgap_case_validation} would have been shortlisted on average. The mean normalized uncertainty penalty differs little between the two selections (0.132 versus 0.136 and 0.123 versus 0.136), so the principal effect of the trust term is to shift the shortlist toward synthesizable chemistry close to known polymers rather than to reduce predictive uncertainty.

The shortlists are robust to the trust-component weights: 200 Dirichlet draws centred on the defaults (concentration 20) retain 95\% and 96\% of the default shortlist on average (minimum 83\% and 87\%), and fully random weights retain 91\% (minimum 83\% and 77\%). Removing a single component retains 83--97\%, with similarity to known polymers the most influential component and uncertainty the least. For the selection weight, $w_{\mathrm{trust}} \leq 0.3$ retains 93--97\% of the default shortlist and $w_{\mathrm{trust}} = 0.5$--0.6 retains 73--87\%, whereas $w_{\mathrm{trust}} = 0.8$ lets trust override Pareto depth (mean layer 2.77 and 1.90; 50\% overlap). The default $w_{\mathrm{trust}} = 0.40$ therefore lies in a regime in which trust refines, but does not override, Pareto ranking, and the conclusions of the case studies do not depend on the precise choice of weights.

Finally, the normalized uncertainty penalty introduced in the Methods section changes the shortlists only marginally relative to the unnormalized form (29 of 30 polymers retained in Case Study 1 and 30 of 30 in Case Study 2, with all ten candidates of Tables~\ref{tab:thermal_case_validation} and~\ref{tab:bandgap_case_validation} retained), but it gives the penalty a usable dynamic range (0.04--0.29 across the Case Study 1 pool, compared with 0.59--0.96 before normalization).

\subsection{Equivalence of AI-Assisted and Manual Discovery}

In addition to the manual multi-objective discovery interface, the platform provides an AI-assisted discovery workflow that enables users to formulate materials-design objectives in natural language. Rather than directly ranking candidates, the language model serves purely as a specification layer that translates user intent into a structured discovery problem consisting of objectives, constraints, and an intent profile; all downstream filtering, ranking, and selection remain governed by the deterministic screening engine described in the Methods section.

To verify that this interpretation layer introduces no discrepancy relative to expert-defined screening, the discovery problem of Case Study 2 was reformulated as a free-text request:

\begin{quote}
\texttt{Find polymers for high-temperature electrical insulation that combine a wide electronic band gap with a high glass-transition temperature, while keeping the density below 1.5 g/cm3.}
\end{quote}

The language-processing module extracted the quantitative design criteria into a structured specification consisting of simultaneous maximization of $E_g^{\mathrm{mono}}$ (the platform's electronic-gap descriptor, to which the requested ``wide electronic band gap'' was mapped) and glass-transition temperature subject to the hard constraint $\rho \leq 1.5$~g~cm$^{-3}$, together with the balanced intent profile, which retains the platform default selection and trust weights. After sanitization and validation, this specification was identical to the manually defined formulation of Case Study 2. Because the discovery engine is deterministic, executing the AI-generated specification reproduced the manual results exactly: the same feasible pool of 960,365 polymers, the same 261 Pareto-layer candidates with identical layer assignments, and the same 30-candidate shortlist with identical ranking order and trust scores, including the five polymers validated in Table~\ref{tab:bandgap_case_validation}. Table~\ref{tab:ai_discovery_example} summarizes this correspondence.

\begin{table}[t!]
\centering
\caption{
Equivalence of the manually defined (Case Study 2) and AI-generated discovery runs for the Case Study 2 problem. The natural-language request was translated into a structured specification that, after validation, matched the manual formulation exactly; the deterministic screening engine therefore produced identical results at every stage.
}
\label{tab:ai_discovery_example}
\small
\begin{tabular}{lll}
\toprule
Component & Manual (Case Study 2) & AI-generated \\
\midrule
Objective 1 & Maximize $E_g^{\mathrm{mono}}$ & Maximize $E_g^{\mathrm{mono}}$ \\
Objective 2 & Maximize $T_g$ & Maximize $T_g$ \\
Hard constraint & $\rho \leq 1.5$ g cm$^{-3}$ & $\rho \leq 1.5$ g cm$^{-3}$ \\
Selection weights ($w_{\mathrm{pareto}}/w_{\mathrm{trust}}$) & 0.60/0.40 & 0.60/0.40 \\
Feasible candidates & 960,365 & 960,365 \\
Pareto-layer candidates & 261 & 261 \\
Shortlist size & 30 & 30 \\
Shortlist overlap & \multicolumn{2}{c}{30/30 (identical candidates, order, and trust scores)} \\
\bottomrule
\end{tabular}
\end{table}

Qualitative emphases expressed in the request are handled through the same auditable mechanism. Wording that stresses reliability, exploration, chemical diversity, strict constraint satisfaction, or restriction to experimentally known polymers is mapped to one of a small set of predefined intent profiles, each corresponding to a documented adjustment of the selection settings rather than an opaque change in ranking behavior; neutral requests, as in the present demonstration, resolve to the balanced profile and reproduce manual screening exactly. Beyond objective generation, the workflow also provides a structured explanation of the generated design strategy, including the rationale behind selected objectives and identified trade-offs, giving users full visibility into the resulting discovery specification without requiring detailed knowledge of the underlying optimization framework. Natural-language interaction therefore makes multi-objective polymer discovery accessible without sacrificing the reproducibility and transparency of the deterministic discovery engine: qualitative design intent and expert-specified screening converge to the same validated candidate set.

\subsection{Literature-Grounded Evidence for the Discovered Polyimide Family}

The shortlists generated in the two discovery case studies are dominated by aromatic polyimide-type repeat units, exemplified by the pyromellitimide backbone of poly[(1,4-phenylenediamine)-\textit{alt}-(pyromellitic dianhydride)], which ranks among the top candidates of both studies. MD validation placed the glass-transition temperatures of these candidates between 650.56 and 759.27~K (Tables~\ref{tab:thermal_case_validation} and~\ref{tab:bandgap_case_validation}), values high enough to warrant independent corroboration. The literature module was therefore applied to the aromatic polyimide family to test whether glass-transition temperatures of this magnitude are supported by published experimental measurements.

The query was issued in material-centered form with polyimide as the material identifier and glass-transition temperature as the free-text target, yielding the assembled search string \texttt{polyimide glass transition temperature} (15 results per source; extraction depth $k=30$; extraction backend \texttt{gpt-oss:latest} through the platform default OpenAI-compatible endpoint). The arXiv, PubMed, and Semantic Scholar sources each returned 15 papers, and title-level deduplication retained 45 unique records, of which the first 30 were passed to full-text acquisition. Automatic PDF retrieval succeeded for one open-access paper; the remaining papers were processed at the bibliographic-record level, with manual PDF upload available as the designed full-text remedy. Contextualized extraction, supplemented by the direct-extraction fallback for 17 papers, produced 20 raw material-property observations from 13 papers: 18 glass-transition records and 2 permittivity records, the latter confirming that staged evidence follows the shared platform property schema rather than a property-specific format.

All 20 observations were parsed, standardized, and validated without loss. Seventeen of the 18 glass-transition records were reported in degrees Celsius and converted under the unit map $\mathcal{S}_{T_g}$; a reported value of 486~$^\circ$C for an ethynyl-crosslinked polyimide composite, for example, was staged as 759.15~K. The additive quality score assigned 8 records to the gold tier and 12 to the silver tier, experimental conditions were captured for 8 records, and all 20 records entered the reviewer queue in pending status.

Table~\ref{tab:literature_case_evidence} lists eight representative records. The staged experimental values span 410.15 to 759.15~K across substituted aromatic polyimides, with the rigid, imide-dense chemistries clustering between 605 and 759~K, including 633.15~K for a Kapton-type polyimide. The MD-validated glass-transition temperatures of the shortlisted candidates therefore lie within the experimentally reported span for this polymer family and coincide with its upper cluster of heavily imidized backbones, providing source-grounded experimental support for the chemistries prioritized by the discovery workflow. Because most records were staged from bibliographic metadata rather than full text, condition capture is sparse, and final acceptance remains a reviewer decision, consistent with the evidence-staging design described in the Methods section.

\begin{table*}[t!]
\centering
\caption{Representative staged literature evidence for the glass-transition temperature of aromatic polyimides from the archived retrieval run of July 13, 2026. Reported values are shown as extracted from the source papers, standardized values follow the property-specific unit map $\mathcal{S}_{T_g}$, and quality tiers follow the additive score $Q_i$. DSC denotes differential scanning calorimetry. All records remain subject to reviewer approval.}
\label{tab:literature_case_evidence}
\scriptsize
\renewcommand{\arraystretch}{1.45}
\setlength{\tabcolsep}{5pt}
\begin{tabular}{lccccc}
\toprule
\textbf{Material} & \textbf{Reported} & \textbf{$T_g$ (K)} & \textbf{Method} & \textbf{Tier} & \textbf{Source} \\
\midrule
Benzoheterocycle polyimide (BIBOPI)& 410~$^\circ$C&683.15& --& Gold&\cite{guo2025benzoheterocycle} \\
Shape-memory polyimide (SMPI)& 205~$^\circ$C& 478.15& --& Gold& \cite{li2021shapememory} \\
Liquid-crystalline poly(ester imide) (LCPEI)& 167~$^\circ$C& 440.15& DSC& Gold& \cite{pan2026benzocyclobutene} \\
Ethynyl-crosslinked polyimide composite (EBPA-2)& 486~$^\circ$C& 759.15& --& Silver& \cite{sun2026hightg} \\
Triarylamine-based wholly aromatic polyimide (PI3M-4M)& 470~$^\circ$C& 743.15& --& Silver&\cite{shao2021triarylamine} \\
6FDA--TFMB/POSS polyimide hybrid& 372~$^\circ$C& 645.15& --& Silver& \cite{lan2019transparent} \\
Kapton-type polyimide& 360~$^\circ$C& 633.15& --& Silver& \cite{wan2025ultrahigh} \\
Silicon-alkyne-functionalized polyimide& 354.6~$^\circ$C& 627.75& --& Silver& \cite{chen2026highperformance} \\
\bottomrule
\end{tabular}
\end{table*}

\section{Discussion}

\subsection{Discovery Across Multiple Polymer Design Domains}

The two discovery case studies, together with the natural-language equivalence demonstration, show that the proposed workflow is not limited to a specific property family or application area. The same discovery engine addressed a lightweight-structural design problem, formulated as simultaneous minimization of density and maximization of glass-transition temperature, and a dielectric-materials design problem, formulated as simultaneous maximization of a monomer HOMO--LUMO-gap proxy for electronic insulation and glass-transition temperature under a density constraint. In each case, the workflow reduced a search space containing more than one million real and virtual polymers to a shortlist of 30 chemically plausible candidates suitable for higher-fidelity evaluation.

Although the targeted property families differ substantially, thermomechanical in the first case and electronic--thermal in the second, both problems were solved using the same underlying sequence of feasibility filtering, exact Pareto ranking, trust-aware prioritization, and diversity-aware selection. The equivalence demonstration further showed that this sequence is reachable through natural language without loss of reproducibility: a free-text request was translated into a validated specification that regenerated the manual results of Case Study 2 exactly.

These results illustrate an important characteristic of the platform: discovery is formulated as a general multi-objective optimization problem rather than as a property-specific workflow. New design problems can therefore be expressed through different combinations of objectives and constraints without modifying the underlying ranking methodology. The resulting flexibility is particularly important in polymer design, where candidate selection is often governed by competing requirements rather than optimization of a single property.

The MD and DFT validation results, interpreted with the known density bias of the MD protocol and the limitations of the gap proxy, further support the usefulness of the workflow as a candidate-prioritization tool. Rather than attempting to replace higher-fidelity simulations, the discovery engine rapidly identifies a small subset of promising candidates from an otherwise intractable search space. This hierarchical strategy enables computational resources to be focused on the most promising polymers while preserving traceability between screening predictions and downstream validation.

A distinguishing feature of the present framework is the integration of discovery, evidence gathering, and simulation preparation within a single workflow. Existing polymer-informatics tools often focus on individual components of the discovery process, such as property prediction, structure generation, workflow automation, or language-based interaction. In contrast, the present platform connects candidate generation, multi-objective ranking, literature support, and simulation handoff through a common data and decision framework. In the present study, this integration was exercised end to end: the polyimide-dominated shortlists produced by the discovery engine were corroborated by staged experimental evidence from the literature module and transferred to simulation-based validation within the same environment. The objective is not to replace existing tools but to reduce the fragmentation that frequently exists between screening, validation, and decision-making stages.

\subsection{Limitations and Future Opportunities}

Several limitations should be acknowledged. First, the trust score is a heuristic combination of similarity to known polymers, internal consistency, predictive uncertainty, and synthesizability. While this ranking strategy improves prioritization of chemically plausible candidates, it should not be interpreted as a calibrated probability of experimental success. Future work may incorporate experimentally validated confidence models or Bayesian decision frameworks to provide more rigorous uncertainty quantification.

Second, both the lookup and prediction layers operate primarily on repeat-unit representations and therefore do not explicitly account for chain architecture, crystallinity, morphology, processing history, or other higher-order structural factors known to influence polymer performance. As a result, the platform should be viewed as a screening and prioritization tool rather than a complete representation of polymer behavior.

Third, the quality of discovery is ultimately limited by the quality of the underlying property datasets and surrogate models. Although the current database integrates experimental, MD, DFT, and group-contribution information, some predicted properties remain associated with substantial uncertainty, particularly in sparsely sampled regions of chemical space. Continued expansion of experimentally validated datasets and multi-fidelity learning strategies will improve the reliability of future predictions.

Fourth, two of the higher-fidelity quantities carry known systematic limitations. The ADEPT MD protocol underestimates experimental densities by about 12\% (median over 414 polymers), so MD densities are interpreted after calibration, and the monomer HOMO--LUMO gap overestimates measured optical gaps by 5--10~eV and cannot capture intermolecular charge-transfer absorption, so it serves only as a relative ranking descriptor. Because the surrogate models were trained partly on data generated with the same MD and DFT protocols, the validation tests surrogate fidelity and protocol consistency rather than providing fully independent physical confirmation. Ensemble spread also cannot reveal erroneous training labels, as the outlier in Case Study 1 illustrates; screening the training data for physically implausible values is a direct route to improvement.

Fifth, the literature workflow currently combines deterministic retrieval with extraction models whose performance depends on document quality and publisher formatting. Reviewer intervention therefore remains necessary for verification of extracted information. Similarly, the AI-assisted discovery workflow simplifies user interaction but does not eliminate the need for scientific judgment when defining meaningful design objectives and constraints.

Finally, the present study validates the discovery workflow through higher-fidelity computational calculations rather than closed-loop experimental realization. Experimental synthesis and characterization of the identified candidates represent an important future direction and would provide the most rigorous assessment of the practical value of the proposed platform.

Despite these limitations, the results demonstrate that integrated discovery workflows can substantially reduce the effort required to navigate large polymer design spaces. By combining multi-source property information, ML predictions, trust-aware screening, literature support, and automated simulation preparation, the platform provides a foundation for more efficient and traceable polymer discovery.

\section{Conclusions}

We presented PolyDiscover, an integrated polymer-discovery platform that combines property lookup, ML prediction, multi-objective screening, literature-assisted evidence gathering, AI-assisted specification generation, generative candidate exploration, and automated simulation handoff within a unified workflow. By consolidating these traditionally disconnected tasks into a single framework, the platform enables traceable progression from candidate exploration to simulation-ready polymer systems while maintaining transparency in the underlying decision-making process.

The capabilities of the platform were demonstrated through two representative discovery case studies spanning lightweight-structural and dielectric-material applications, complemented by a demonstration that natural-language specification exactly reproduces manually defined screening. In each case study, the workflow successfully reduced a search space containing more than one million real and virtual polymers to a small set of chemically plausible candidates using feasibility filtering, Pareto ranking, trust-aware prioritization, and diversity-aware selection. An ablation of the trust term showed that it shifts about one fifth of each shortlist toward synthesizable chemistry close to known polymers without sacrificing Pareto depth, and that the shortlists are robust to the choice of weights. Subsequent higher-fidelity validation further illustrated the utility of the framework as a candidate-prioritization tool, enabling computational resources to be focused on the most promising regions of polymer chemical space, while staged literature evidence provided independent experimental support for the polyimide-dominated shortlists.

In combination, these components form a foundation for more efficient and reproducible polymer design by connecting data resources, predictive models, and physics-based simulations within a common computational ecosystem. Future development will focus on expanding multi-fidelity datasets, improving uncertainty quantification, incorporating additional property domains, and enabling closed-loop experimental validation of discovered candidates.

\section*{Author Contributions}

S.A.: conceptualization, methodology, software, investigation, data curation, writing. G.X.: software, investigation. Y.L.: investigation, validation. R.Z.: investigation, validation. M. J. and T. L. supervised the research and contributed to the conceptual development of the study. All authors contributed to writing and reviewing the manuscript.

\section*{Conflicts of Interest}
There are no conflicts to declare.

\section*{Acknowledgements}
This work was supported by the Lucy Family Institute for Data \& Society Postdoctoral Fellowship and National Science Foundation grants 2332270 and 2102592. The authors also acknowledge the Center for Research Computing at the University of Notre Dame for providing the necessary computing resources.

\section*{Data and Code Availability}
The PolyDiscover platform is publicly accessible at \url{https://huggingface.co/spaces/sobinalosious92/POLYMER-PROPERTY}, where the complete source code is maintained. The processed datasets and scripts required to reproduce the reported results, including the discovery specifications and candidate shortlists for both case studies and the scripts used for the uncertainty normalization, trust ablation, weight-sensitivity, MD-density calibration, and gap-proxy analyses, are provided in the same repository. The underlying ADEPT--PolyGraphMT simulation workflow, predictive models, and training datasets are available as described in the companion publication.\cite{alosious2026adept}

\bibliography{reference}

\end{document}